\documentclass[prb,showpacs,superscriptaddress,twocolumn,notitlepage]{revtex4-1}

\usepackage{amsmath}
\usepackage{amssymb}
\usepackage{amsfonts}
\usepackage{graphicx}
\usepackage{multirow}
\usepackage{chngcntr}
\usepackage{bm}

\usepackage{color}
\usepackage{ulem}
\usepackage[usenames,dvipsnames]{xcolor}

\usepackage{natbib}
\usepackage[final=true]{hyperref}
\hypersetup{
    colorlinks=true,
    linkcolor=Maroon,
    filecolor=magenta,
    urlcolor=cyan,
    citecolor=PineGreen,
}

\newcommand{\be}{\begin{eqnarray}}
\newcommand{\ee}{\end{eqnarray}}

\begin{document}


\title{Localized and extended collective optical phonon modes in regular and random arrays of contacting nanoparticles: escape from phonon confinement}


\author{S.~V.~Koniakhin$^*$}
\affiliation{Center for Theoretical Physics of Complex Systems, Institute for Basic Science (IBS), Daejeon 34126, Republic of Korea}
\affiliation{Basic Science Program, Korea University of Science and Technology (UST), Daejeon 34113, Korea}
\email{kon@ibs.re.kr}

\author{O.~I.~Utesov}
\affiliation{Center for Theoretical Physics of Complex Systems, Institute for Basic Science (IBS), Daejeon 34126, Republic of Korea}
\affiliation{The Faculty of Physics of St. Petersburg State University, Ulyanovskaya 1, St. Petersburg 198504, Russia}

\author{A.~G.~Yashenkin}
\affiliation{The Faculty of Physics of St. Petersburg State University, Ulyanovskaya 1, St. Petersburg 198504, Russia}
\affiliation{Petersburg Nuclear Physics Institute NRC ``Kurchatov Institute'', Gatchina 188300, Russia}

\begin{abstract}



In the present paper, we utilize the coupled-oscillator model describing the hybridization of optical phonons in touching and/or overlapping particles in order to study the Raman spectra of nanoparticles organized into various types of regular and random arrays including nanosolids, porous media, and agglomerates with tightly bonded particles. For the nanocrystal solids, we demonstrate that the ratio of the size variance to the coupling strength allows us to judge the character (localized or propagating) of the optical phonon modes which left the particles of their origin and spread throughout an array. The relation between the shift and the broadening of the Raman peak and the coupling strength and the disorder is established for nanocrystal solids, agglomerates, and porous media providing us with information about the array structure, the structure of its constituents, and the properties of optical phonons.

\end{abstract}

\maketitle

\section{Introduction}

Nanotechnology revolutionized the way we treat materials, opening up exciting possibilities for numerous scientific and technological advancements. Among the currently studied nanoobjects, significant progress is achieved in manufacturing, characterization, and utilization of nanoparticles~\cite{khan2019nanoparticles,mochalin2020properties, Mitchell2021}, which not only can inherit the properties of basic bulk materials but also possess a distinct set of physical and chemical features considerably different from the bulk ones. Such features as large surface-to-bulk ratio and quantum confinement effects play an important role in nanoparticles. It is pertinent to also mention several useful properties, including enhanced catalytic activity~\cite{astruc2020introduction,chen2021atomically}, tunable optical and electronic characteristics~\cite{terna2021future,chernikov2023tunable,nunn2023optical}, bio-compatibility \cite{bao2010biosynthesis,fusco2020impact,leung2023versatile}, which attracted significant attention and propelled the rapid growth of nanoparticle-based research and development. By tailoring the size, shape, and composition of nanoparticles, researchers can precisely engineer their properties to meet specific requirements for diverse applications. Among various types of nanostructures, a significant role is played by the carbon nanostructures~\cite{das2023polymer,nyholm2023functionalized} and, specifically, the diamond nanoparticles obtained using various synthesis methods~\cite{mochalin2012properties,wood2022long}.

Along with benefiting from the direct usage of nanoparticles, nanocrystals, and quantum dots for various technological and scientific applications, they emerged as a key building block for materials of the higher levels of organization. Such materials are the quantum-dot molecules~\cite{panfil2019electronic,cui2019colloidal,koley2021coupled,cui2021neck,cui2021semiconductor}, nanocrystal solids~\cite{bozyigit2014electrical,zhao2021enhanced} and networks~\cite{jang2015solution,evers2015high,whitham2016charge,chen2016metal,reich2016exciton}. In addition, the porous materials~\cite{yang1994study,alfaro2008theory,alfaro2011raman,valtchev2013porous,kosovic2014phonon} with certain geometry can be considered as networks of coupled nanoparticles.

Furthermore, the nanoparticle agglomerates whose formation is inherent to the specific manufacturing techniques can also serve as an example of coupled nanoparticles arrays. The diamond nanoparticles of detonation synthesis can serve as an example. Primary crystallites of approx 3-5~nm in size~\cite{raty2003ultradispersity,ozerin2008x,koniakhin2015molecular,koniakhin2018ultracentrifugation,trofimuk2018effective} in hydrosols and dried powders are organized into stable agglomerates\cite{baidakova1999ultradisperse,williams2007enhanced,osawa2008monodisperse,avdeev2009aggregate,tomchuk2015structural,dideikin2017rehybridization} with fractal structure \cite{avdeev2009aggregate,tomchuk2015structural,koniakhin2020evidence}. The deagglomeration of particles is quite involved and requires chemical purification and ultrasonication in hard conditions, leading to the establishment of covalent bonding between the particles via the carbon phase undergoing the amorphization~\cite{dideikin2017rehybridization}.

Raman spectroscopy (RS), a potent technique, finds extensive application in the characterization of contemporary nanostructured materials such as nanoparticles, nanorods, and two-dimensional nanostructures. It provides an accurate energy profile of material-specific excitations like phonons~\cite{korepanov2017carbon,vinogradov2018structure,korepanov2020localized}, magnons~\cite{qiu2022two}, and excitons~\cite{zucker1987resonant,brewster2009exciton}. Currently, the Raman spectra measurements serve as the standard procedure for characterizing carbon materials and nanomaterials designed for various applications~\cite{ferrari2004raman,osswald2009phonon,kumar2012raman,korepanov2017carbon,vinogradov2018structure,lee2019raman,ourBench,kashyap2021comparative}. By offering ease of implementation, nondestructive analysis, and a wealth of versatile data, Raman spectroscopy plays a pivotal role in advancing nanotechnology and material science. Importantly, when the nanoparticles are concerned, the theoretical basis of RS deals usually with single-particle properties only; the resulting spectra are evaluated as integrated over {\it independent and non-interfering} particles of an array.

Recently, the theory of optical phonon mode hybridization has been developed for two and several nanoparticles in contact and the corresponding effects on Raman spectra have been studied in Ref.~\cite{koniakhin2023coupled}. This theory predicts the vibrational properties of a nanoparticle dimer (single quantum dot molecule). It has been demonstrated that optical phonons in such a system behave qualitatively similar to the system of two coupled harmonic oscillators. For a quantitative description, the Coupled Oscillators Model (COM) has been proposed, and formulated as an eigenproblem for $2 \times 2$ matrix with certain coefficients. Despite COM simplicity, it has been shown to reproduce the results of more involved approaches, namely the atomistic dynamical matrix method (DMM)~\cite{ourDMM} and the continuous Euclidean metric Klein-Fock-Gordon equation (EKFG)~\cite{ourEKFG}.

In the present study, we extend the COM approach onto several physically relevant many-particle ensembles and explore the effects of particle-particle contacts on optical phonon hybridization. We investigate how the abovementioned phenomena manifest themselves in the Raman spectra of the corresponding arrays.  In particular, we study
the regular 3D arrays of nanoparticles  (ordered and disordered) as a model for nanosolids and porous media.
Furthermore, we evaluate the inverse participation ratio (IPR) as a function of phonon mode hybridization and disorder.
We obtain (both analytically and numerically) the scale of phonon localization, and attribute our problem as an Anderson transition~\cite{anderson1958absence,abrahams2010}.
We also extend our theory onto 2D, 1D arrays, and tight agglomerates.

Since below we discuss the propagation/localization of vibrational modes on both regular and disordered arrays we would like to emphasize
the type of optical phonons we are dealing with. It is precisely the {\it intrinsic} modes of individual particles which are usually treated as a subject of the phonon confinement~\cite{richter1981one,campbell1986effects}. In our approach, they acquire an opportunity to escape the maternal particles and to travel throughout an array due to the hybridization of modes from neighboring particles that occurs due to their contacts (c.f. hybridization of Lamb modes described in Ref. \cite{jansen2023nanocrystal}). These modes should be contrasted with the collective vibrational modes (superlattice phonons) of regular enough arrays whose eigenfrequencies can be obtained from, e.g., the ball-spring model with suitable interparticle interaction \cite{yazdani2019nanocrystal,jansen2019phonon}. The latter are not a subject of the present study.

The rest of the paper is organized as follows. In section~\ref{sec_formulation} we formulate the generalization of the COM approach for the case of a many-particle array/network. In section~\ref{sec_results} we describe the results for various types of nanoparticle arrays. In subsection~\ref{subsec_sec_porous} we consider three-dimensional regular arrays of identical particles and establish the correspondence between these systems and the porous media. In subsection~\ref{subsec_3D} we study a 3D regular array of nanoparticles varying in size. The criterion of optical phonons localization in such a system is formulated depending on the size variance and the particle overlapping parameter. Subsection~\ref{subsec_2D} describes the model cases of the 2D array and 1D chain. In subsection~\ref{subsec_agglomerates} we consider the nanoparticle agglomerates with the spatial structure governed by cluster-cluster and cluster-particle aggregation processes. Section \ref{sec_concl} is devoted to the analysis of the obtained results and presents the main conclusions of the conducted study.




\section{Coupled-oscillator model for optical phonons in many-particle systems}
\label{sec_formulation}

In high-symmetry lattices,  (e.g., in  diamond-type lattices), the energy spectrum of long-wavelength optical phonons near the Brillouin zone center is parabolic
\be \label{spec1}
  \omega(q) \approx \omega_0 - \alpha q^2
\ee
in accordance with Keating model~\cite{keating1966}. Evidently, it is consistent with the widely used form of dispersion \mbox{$\omega(\Tilde{q}) \approx A + B\cos(\Tilde{q})$}~\cite{ager1991spatially,yoshikawa1995raman,chaigneau2012laser}. The maximal optical phonon frequency for a diamond is $\omega_0 \equiv A + B \approx 1333$~cm$^{-1}$.

In nanoparticles, the optical vibration frequencies can be found, e.g., by direct diagonalization of the system dynamical matrix~\cite{ourDMM} or using the continuous EKFG approach~\cite{ourEKFG}. The latter implies the solution of the Laplace equation with the Dirichlet boundary condition
\be \label{EKFG}
  \Delta Y + q^2 Y = 0, \quad \quad Y|_{\partial \Omega} =0
\ee
on the connected manifold $\Omega$ corresponding to the nanoparticle shape. Here $Y$ represents the optical mode envelope function. For simple shapes, e.g., for sphere, cylinder, or cube, Eq.~\eqref{EKFG} can be easily solved analytically and the set of eigenvalues $q^{2}$ gives the optical phonon spectrum. For instance, in the case of spherical particles, the phonon wave functions are combinations of spherical Bessel functions and spherical harmonics (see Ref.~\cite{ourEKFG}). The lowest eigenvalue reads $q_1 = 2 \pi/L$  for the $s$-wave state in a nanoparticle of the size (i.e., diameter) $L$. The corresponding wave function is the most symmetrical one among all modes and gives the strongest Raman signal $I \propto |\int_{\Omega} Y d{\bf r}|^2$ from a single particle~\cite{ourEKFG}. Due to the dispersion law with negative mass, this mode is highest in energy.

However, when the manifold is a complex one, the direct solution of Eq.~\eqref{EKFG} is more involved. The calculations remain reasonable in size if we are dealing with non-interacting nanoparticles, but when we consider the effect of their cross-talks (this is what we do in the present study) the calculation time increases dramatically. This is indeed the case for systems that consist of many structural units, e.g., nanocrystal solids, porous media, and tight nanoparticle agglomerates. In order to overcome this problem, recently the coupled-oscillator model (COM) was proposed~\cite{koniakhin2023coupled}. It allows to reformulate the problem of the optical phonon spectrum of several nanoparticles in contact as a simple eigenvalue problem for a certain matrix. For example, for two interpenetrated particles the corresponding matrix reads
\be \label{eq_coupledOsc_ss}
  H = \left(
  \begin{array}{cc}
   q_1^2 - \Delta_{1} & - C_{12} \\
   - C_{21} & q_2^2 - \Delta_{2} \\
  \end{array}
\right).
\ee
This matrix, which can be referred to as the problem Hamiltonian, describes the coupling and hybridization of two highest-in-energy optical modes belonging to two particles in contact. Here $q^2_{1,2}$ are the known smallest eigenvalues of the two particles when they are isolated, $\Delta_{1,2}$ are the on-site corrections due to the change of the volume accessible for the mode resulting in consequent slight relaxation of size-quantization effect, and $C_{12} = C_{21}$ are the inter-mode coupling constants. The crucial parameter here is the particles' intersection volume $V_{12}$ which is assumed to be much smaller than the volumes of both particles $V_{1,2}$. For a spherical shape, it is given by the following equation:
\be
  \label{eq_V12}
  V_{12}(R_1,R_2,\delta R) \approx \frac{\pi \delta R^2 R_1 R_2}{R_1+R_2},
\ee
where $R_{1,2}$ are the particles radii and $\delta R$ is the penetration length (overlapping parameter). Now we can present explicit equations for the model parameters. For the on-site corrections, one has
\be
  \Delta_{1,2} = q^2_{1,2} f(V_{12}/V_{1,2}), \quad f(x)=x-(x/0.425)^2,
\ee
whereas the coupling constant reads
\be
  C_{12} = \frac{\sqrt{ q^2_{1} q^2_{2}}V_{12}}{\sqrt{V_1V_2}}.
\ee
In our previous study COM was shown to accurately reproduce the results of the EKFG approach and, on a semi-quantitative level, the ones of an exact dynamical matrix method.

\begin{figure}
  \includegraphics[width=8.cm]{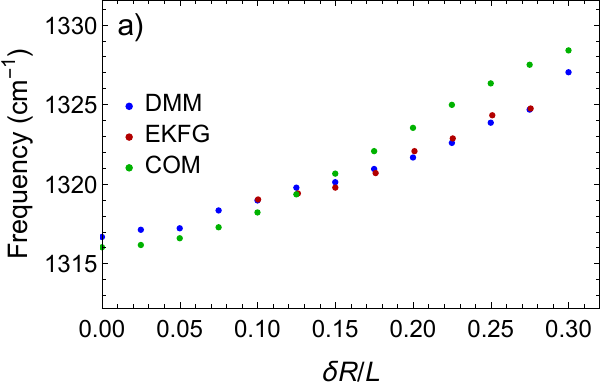} \\
  \includegraphics[width=8.cm]{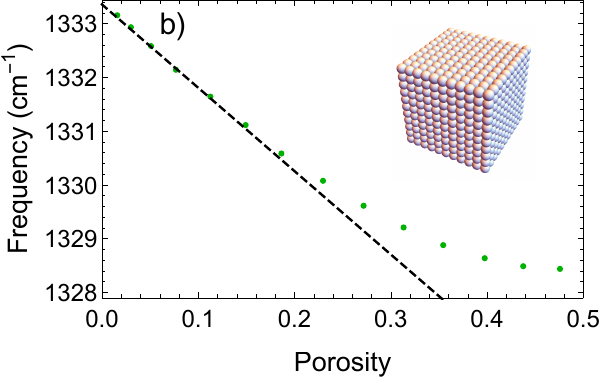}
  \caption{Panel (a). The highest optical phonon frequency of $3\times3\times3$ regular array of particles arranged into the simple cubic lattice obtained within DMM, EKFG, and COM approaches as a function of the penetration length. The case of 1.6~nm spherical nanodiamond is considered. Panel (b). The highest optical phonon frequencies of $11\times11\times11$ regular array of 3~nm diamond particles obtained using the COM approach. The limiting value at zero porosity 1333.3~cm$^{-1}$ matches the value of the bulk diamond optical phonon frequency. The simulated structure is depicted in the inset.}
\label{fig_1_regular}
\end{figure}

Importantly, COM allows for the simple generalization onto many particle arrays. Let the corresponding number of particles be $N$. Then, one should accurately calculate the on-site corrections, which for $i$-th particle reads
\be
  \Delta_{i} = q^2_{i} \sum_j f(V_{ij}/V_i).
\ee
Next, the coupling constants are
\be
  C_{ij} = \frac{\sqrt{ q^2_{i} q^2_{j}}V_{ij}}{\sqrt{V_iV_j}}.
\ee
So, the $N \times N$ matrix problem has the following form:
\be \label{ham1}
  H = \begin{bmatrix}
    & \ddots & \vdots & &\\
    & \dots&q_i^2 - \Delta_i & \dots  & -C_{ij} & \dots\\
    & &\vdots & \ddots & & \\
    &  &-C_{ij} &   & \ddots  &  \\
    &  &\vdots &   &  & \ddots
    \end{bmatrix}
\ee
Finally, solving the eigenproblem
\be
  H \psi = \lambda \psi,
\ee
one obtains a set of eigenvalues $\lambda$ [which should be plugged into Eq.~\eqref{spec1} instead of $q^2$ in order to get the phonon energies] and eigenvectors $\psi^{(n)}$, which indicate the probabilities to find an optical phonon at a certain particle. Importantly, the latter can be also utilized for Raman intensity calculations. For a particular mode $\nu$ it reads
\be \label{ramanI}
  I_\nu = \left| \sum^N_{i=1} \sqrt{V_i} \, \psi_i^{(\nu)} \right|^2,
\ee
where the summation includes all the particles in the ensemble (cf. Ref.~\cite{ourEKFG}). The Raman spectrum of an array can be derived using the equation
\be 
I(\omega) = \sum_{\nu} I_\nu \delta (\omega-\omega_\nu).
\ee
The delta function here can be replaced by a Lorentzian profile with an appropriate intrinsic or experimental broadening.

Note that the matrix~\eqref{ham1} corresponds to the Anderson-like model~\cite{anderson1958absence} with the on-site disorder and variable hopping elements. Thus, depending on the model parameters, the eigenmodes, in our case, the optical phonons, can be either localized or extended (propagating). We show that this phenomenon plays a crucial role when the Raman spectrum of the ensemble is concerned. For the quantitative characterization of the localization properties, we use the inverse participation ratio (IPR) defined as follows
\be \label{IPR}
  \mathrm{IPR}(\nu) = \sum_i \left|\psi_i^{(\nu)} \right|^4
\ee
and assume function normalization to unity.

For extended states $\mathrm{IPR} \sim 1/N^\alpha$ with some $\alpha > 0$, whereas for localized ones $\mathrm{IPR}$ saturates to some constant upon $N$ growth ($\mathrm{IPR} \sim N^0 \equiv \textrm{const}$). Parameter $\alpha$ indicates the fractal dimension of the corresponding eigenstate. Its flow upon the decimation procedure~\cite{aoki1980} allows to judge the localization properties of various models including random regular graphs~\cite{vanoni2023renormalization}.

\section{Results for ordered and disordered systems}
\label{sec_results}

\subsection{Three-dimensional nanocrystal solid}

\label{subsec_sec_porous}

In this subsection, we test COM applicability using a relatively simple problem of optical phonons in a nanocrystal solid. The latter is assumed to be made of identical particles arranged in a simple cubic lattice. Evidently, in this case, all the modes will be propagating, i.e., extended over the whole system since there is no disorder in the model. One can also expect that the mode with the highest frequency would have $\omega \to \omega_0$ (BZ center optical phonon frequency, 1333~cm$^{-1}$ for Diamond) when the overlap among the particles increases. The latter can be also discussed in terms of decreasing porosity. Hereinafter, we shall use the notion of ``porosity'' in order to quantify the relative measure of free space inside the sample.

We begin our analysis with a comparison between the results of effective COM and more involved  DMM and EKFG approaches. For this purpose, we consider a $3\times3\times3$ system made of $L=1.6$~nm diamond nanoparticles. Then we shall vary the penetration depth $\delta R$ keeping track of the optical mode with the highest frequency. Importantly, the latter has the dominant Raman intensity among all the modes being the most symmetric one. Our findings are presented in Fig.~\ref{fig_1_regular}(a), where one can see a good agreement among all three approaches. Notice that for a regular array on a simple cubic lattice $\delta R = L - a$, where $a$ is the lattice parameter (the shortest distance between the particle centers).
\begin{figure}
  \includegraphics[width=7.cm]{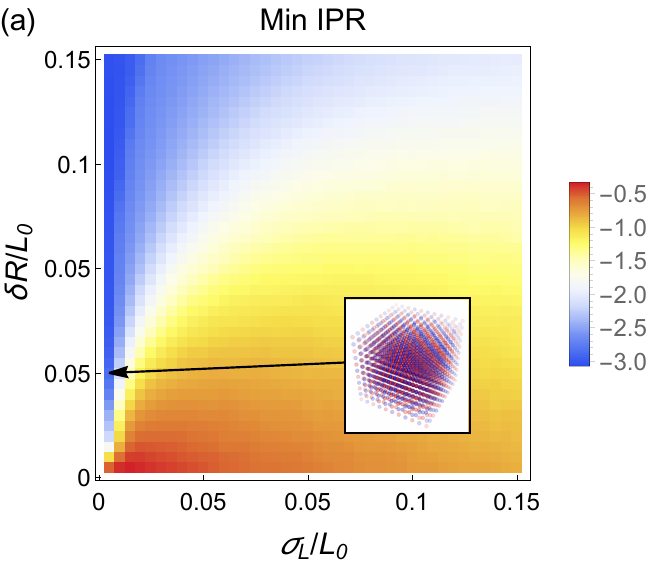} \\
   \includegraphics[width=7.cm]{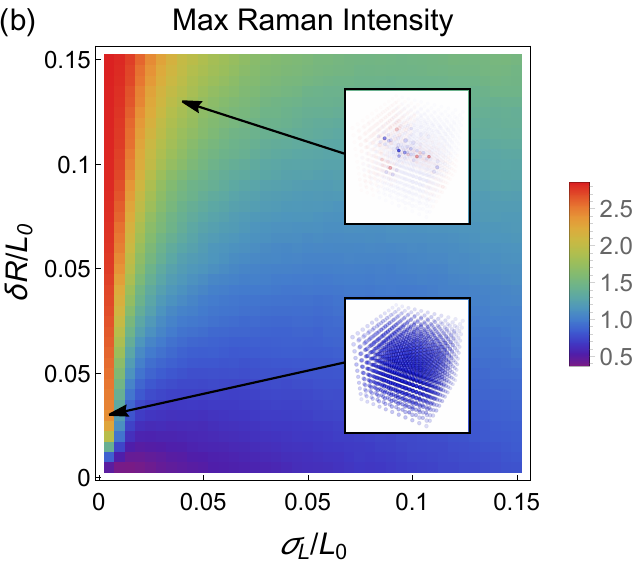} \\
   \includegraphics[width=7.cm]{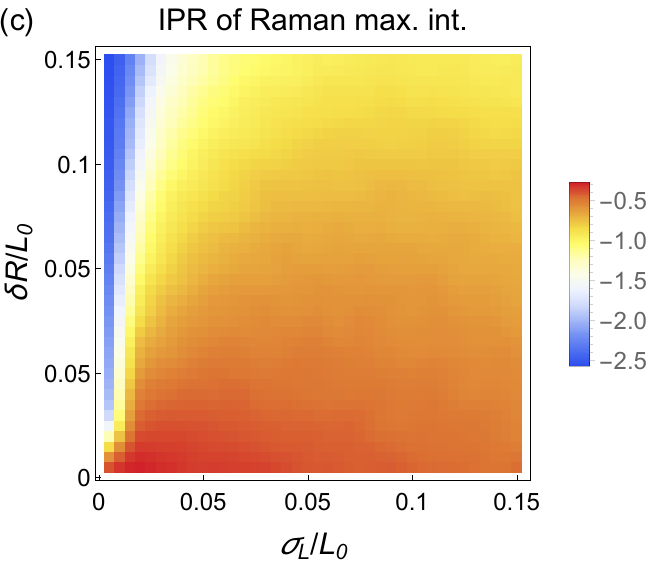}
   \caption{Localization of optical phonons in a 3D array of nanoparticles. Panel (a). The log of the inverse participation ratio of the most delocalized eigenfunction versus the typical size variance $\sigma_L/L_0$ and penetration length $\delta R/L_0$. (b) The same for the maximal Raman intensity of a single mode. Panel (c) depicts the IPR of the mode that has maximal Raman intensity. The insets show the real-space amplitudes of the corresponding wave functions.}
\label{fig_3D_phase_diag}
\end{figure}
Next, we turn to the more complicated case of $11\times11\times11$ 3~nm particles array, which can be hardly analyzed using numerical DMM or even EKFG. Its total size is \textit{approx.} 30~nm ( depending on the particle overlapping length $\delta R$) which is almost a bulk material from the point of view of the phonon confinement. Using COM, we examine the maximal optical phonon frequency as a function of the $\delta R/L$ parameter. The maximal porosity of the lattice of slightly touching spheres ($\delta R=0$) is \textit{approx.} $0.48$. The value $\delta R/L=0.3$ yields the porosity less than 0.03, so the particles are essentially overlapped. Fig.~\ref{fig_1_regular}(b) shows that the maximal optical phonon frequency almost reaches $\omega_0$ in a small porosity limit, which is quite expected since this limit is close to the bulk case. In another limiting case of weakly penetrated spheres, the frequency tends to the maximal frequency of a single particle, which is $1327$~cm$^{-1}$ according to the EKFG model.

To conclude this subsection, we demonstrated that COM works reasonably well for hybridized optical vibrations in many-particle systems. Below we shall utilize this knowledge for disordered arrays.


\subsection{Regular array with disorder}

\label{subsec_3D}

Now let us take the model from the previous subsection and ``spoil'' it. The simplest yet natural way to introduce disorder is to introduce the size distribution for particles arranged in the nanocrystal solid, say, in the Gaussian form:
\begin{equation} \label{dis1}
    P(L) \propto \exp[-(L-L_0)^2/2\sigma_L^2],
\end{equation}
where $L_0$ stands for the average nanoparticle size and $\sigma_L$ shows the typical deviation from it. Keeping the distance between the centers of neighboring particles intact, one can see that the variety in particle sizes results in disorder in both diagonal and off-diagonal matrix elements in Hamiltonian~\eqref{ham1}. In order to illustrate this issue,  we consider an $11\times11\times11$ array of nanoparticles arranged in a simple cubic lattice with the parameter $a$. Due to the particle size variation~\eqref{dis1}, the ``bare'' penetration depth $\delta R_0 = L_0 - a$ also varies in the present case.

Before presenting our numerical results, let us discuss what can we expect from the model under investigation. Considering the case of moderate $\sigma_L \ll L_0$, we see that the main contribution to disorder in diagonal elements is due to $q^2_i$ variation which can be estimated as
\be
  \delta q^2_i \sim \frac{\sigma_L}{L^3_0}.
\ee
Importantly, the changes in other quantities that are present in the Hamiltonian~\eqref{ham1} are much smaller due to the smallness of the relative particle intersection volume. Taking into account this finding, we arrive at the famous Anderson model with the {\it diagonal} disorder. According to the Thouless Percolation-theory arguments~\cite{thouless1974electrons}, in order to judge the extended/localized character of excitation behavior in this model, one should compare the average deviation in the diagonal matrix elements and the hopping (off-diagonal elements) magnitude. In our case, if $\langle \delta q^2_i \rangle \ll \langle C_{ij} \rangle$ then the eigenstates should be extended, whereas for $\langle \delta q^2_i \rangle \gg \langle C_{ij} \rangle$ the states are localized. The boundary between the two regimes can be estimated as
\be \label{deloc}
  \sigma_L \sim \frac{\delta R^2_0}{L_0}.
\ee

\begin{figure}
  \includegraphics[width=8.cm]{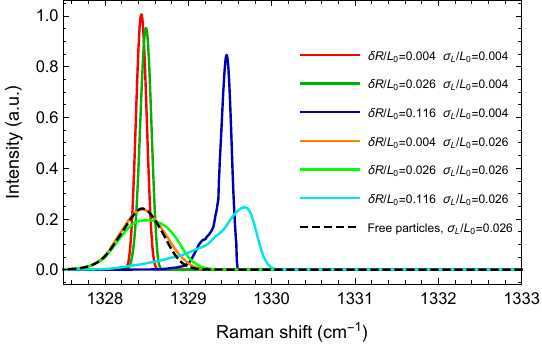} \\
   \caption{Raman spectra for various values of the size distribution variance $\sigma_L/L_0$ and the overlapping parameter $\delta R/L_0$.}
\label{fig_3D_Raman_phase_diag}
\end{figure}

Let us turn to the results of our numerical simulations which were obtained after averaging over 30 disorder realizations. In Fig.~\ref{fig_3D_phase_diag}(a) we plotted the logarithm of the IPR for the mode with minimal IPR, i.e., for the maximally delocalized state. Evidently, one can see the boundary between the regimes of localized and extended states. Moreover, it has a good correspondence to the qualitative estimate~\eqref{deloc}. Thus we indeed observe the Anderson-like transition for optical phonons in our problem. Importantly, the mode with minimal IPR turns out to be the most antisymmetric one with respect to the amplitude of $\psi$ on neighboring particles. Thus, it is characterized by the largest eigenvalue and smallest frequency. Moreover, its Raman intensity is negligible [see Eq.~\eqref{ramanI}].

When the Raman {\it active } modes are addressed, we observe that the ones with the highest intensity have the most symmetric wave functions $\psi$ and the highest frequency. In the delocalized regime, there exists a single mode with dominating intensity as shown in the inset in Fig.~\ref{fig_3D_phase_diag}(b). In contrast, when the states are localized, there are several modes with comparable contributions to the RS. This results in a {\it broader} Raman peaks (see Fig.~\ref{fig_3D_Raman_phase_diag} and discussion below). Fig.~\ref{fig_3D_phase_diag}(c), where the IPR of the mode with the maximal Raman intensity is shown, supports our main claim that the localization properties of the excitations are crucial for Raman spectra understanding. Furthermore, we illustrate it with the direct calculation of the specimen RS for six various parameter sets corresponding to various regimes, see Fig. \ref{fig_3D_Raman_phase_diag}. When the states are extended, the RS has a sharp peak with relatively small linewidth. However, when the disorder is sufficient to localize the eigenmodes, the signal is essentially broadened. In the case of weak coupling, it becomes close to the one, which we can expect for an ensemble of isolated particles with the same size distribution~\eqref{dis1} (see Fig. \ref{fig_3D_Raman_phase_diag}, orange, green, and dashed-black curves). When the coupling (or, equivalently, the average penetration depth) becomes larger, the Raman signal shifts towards higher frequencies as shown by the cyan curve in  Fig.~\ref{fig_3D_Raman_phase_diag}. This phenomenon is associated with the longer localization length in this case so the Raman active modes are mostly localized within several particles and hence their frequencies are higher than those of isolated particles. Noteworthy, as one can see from Fig.~\ref{fig_3D_phase_diag}, maximal Raman intensity expressed by Eq.~\eqref{ramanI} can serve as a characteristic of modes localization along with IPR.

\begin{figure}
  \includegraphics[width=8.cm]{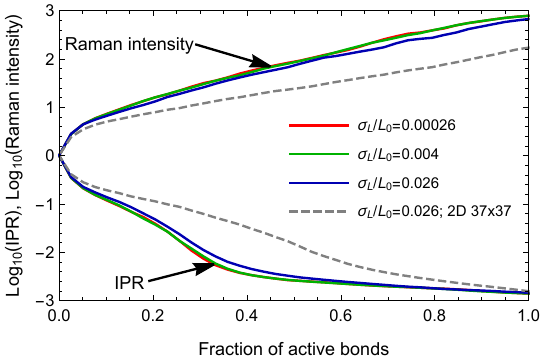}
   \caption{Percolation scenario in the system of $11\times11\times11$ cubic nanoparticle array. The overlap $\delta R/L_0=0.25$ is fixed. The solid curves depict the IPR and the Raman intensity (in units of Raman intensity of a single particle) as a function of the fraction of active bonds in the system.  The dashed gray line illustrates the same effects for a 2D square lattice with 37x37 $\approx$ 11x11x11 sites.}
\label{fig_perc}
\end{figure}

Finally, in order to make a connection with the percolation theory~\cite{nakayama1994dynamical} even more transparent, we consider the same problem but with a certain amount of couplings which we manually switch off. Then, we study the dependence of the maximal Raman intensity of a single mode and minimal IPR among all the modes on the fraction of active bonds $x$. Fig.~\ref{fig_perc} summarizes the corresponding findings in the regime of strong coupling ($\delta R/L_0 = 0.25$). One can see, that in this case, the size distribution function broadening is irrelevant because $\sigma_L \ll L_0$, but the fraction of active bonds $x$ is the crucial parameter. Importantly, the percolation transition occurs at $x_c=0.25$. At $x>x_c$ we indeed observe extended states with small IPR and large Raman intensity, whereas at $x<x_c$ the modes are localized and their intensities are relatively small. Moreover, the dependencies here are found to be in semi-quantitative agreement with the results shown in Fig.~\ref{fig_3D_phase_diag}(b) and (c).




\subsection{The case of 2D and 1D arrays.}

\label{subsec_2D}

After discussing the main properties of the model and establishing its close relation to the 3D percolation problem, let us examine its behavior in lower dimensions investigating the square particle lattice and the linear particle chain.

\begin{figure}
  \includegraphics[width=7.cm]{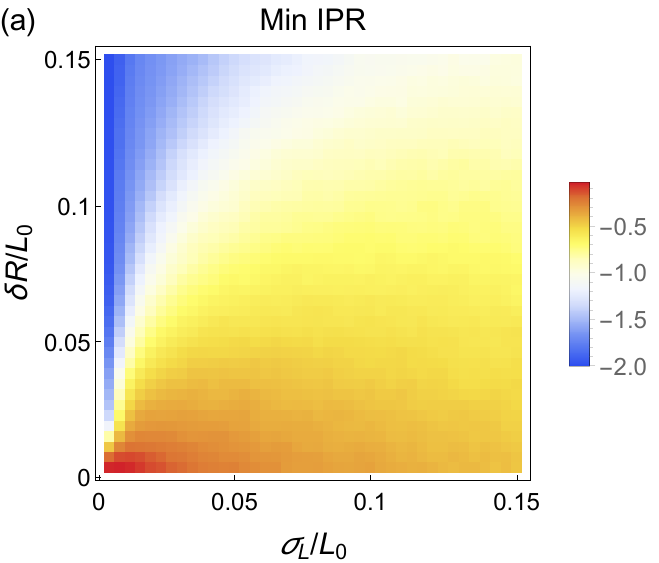} \\
   \includegraphics[width=7.cm]{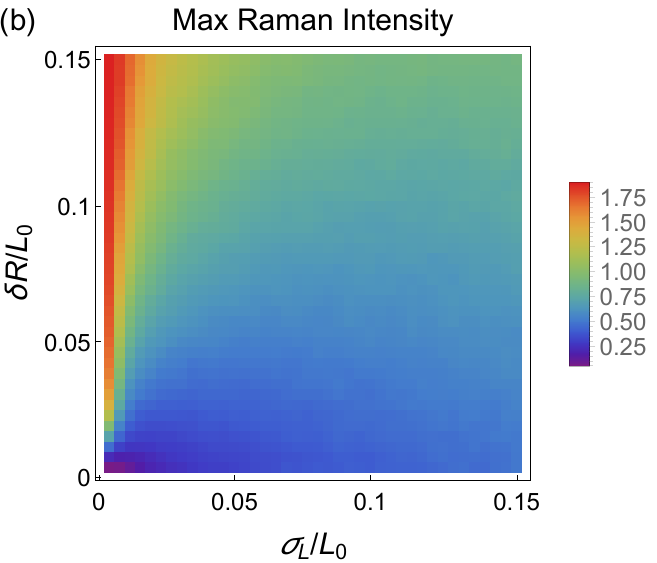} \\
   \caption{Localization of optical phonons in a 2D nanocrystal solid (array of nanoparticles) arranged in a square lattice. Panel (a). The log of inverse participation ratio of the most delocalized eigenfunction versus the typical size variance  $\sigma_L$ and the overlap length $\delta R$. Panel (b). The maximal Raman intensity of a single mode in the same coordinates.}
\label{fig_2D_phase_diag}
\end{figure}

It is well-known that in two dimensions the percolation transition is qualitatively the same as in 3D (see the dashed line in Fig.~\ref{fig_perc}). An important difference is that $x_c = 0.5$ in this case; another difference in values of critical exponents is not important for our crude treatment. Therefore, we can expect a smaller coefficient of proportionality in Eq.~\eqref{deloc} for the localization boundary. This was indeed what we observed in our numerics for $11 \times 11$ lattice, see Fig.~\ref{fig_2D_phase_diag}. Evidently, in order to observe 2D extended states at given $\sigma_L$ one should involve larger $\delta R$ (cf. Fig.~\ref{fig_3D_phase_diag}). Nevertheless, on the qualitative level, the properties of the 2D model remain the same as discussed above for 3D. 

It is pertinent to note that it is well-known that in 2D even for weak disorder all states are localized~\cite{abrahams2010}. However, the localization length is exponentially large and we cannot observe it in our calculations. Instead, we discuss strong localization which is qualitatively similar to 3D.

\begin{figure}
  \includegraphics[width=7.cm]{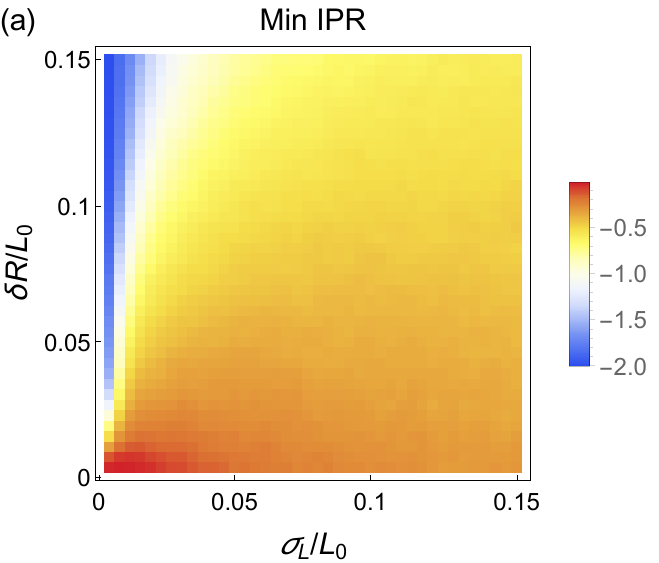} \\
   \includegraphics[width=7.cm]{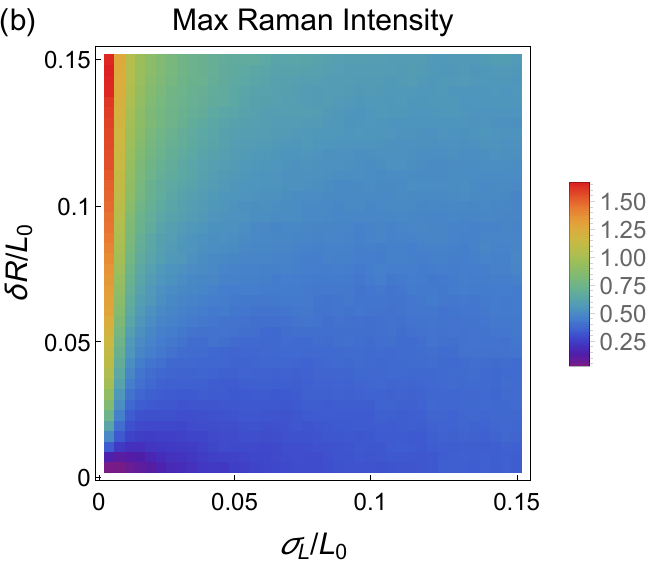} \\
   \caption{Localization of optical phonons in 1D chain. Panel (a). The log of the inverse participation ratio of the most delocalized eigenfunction versus the typical size variance $\sigma_L$ and the typical overlap parameter $\delta R$. (b) The same for the maximal Raman intensity of a single mode.}
\label{fig_1D_phase_diag}
\end{figure}

In one spatial dimension, the physics is essentially different. For infinitely large chains the percolation threshold is exactly $x_c=0$ since every ``inactive'' bond breaks the system into two separate pieces. However, since we consider the system with finite size $N$, the physics is more tricky. An additional parameter -- localization length $\xi$ -- comes into play. If $\xi \gtrsim N$ then, the optical phonons are in the extended regime. In the opposite case of $\xi < N$ the localization takes place.

For the simple estimate of  $\xi$ let us assume that $\delta R^2_0 \gg L_0 \sigma_L$ (weak disorder) and calculate the probability of finding a bond with $|q^2_i - q^2_{i+1}| > C_{i,i+1}$. As a result, without numerical coefficients, we obtain
\be
  \xi \sim \dfrac{\delta R^2_0}{L_0 \sigma_L}  \exp{\left(\dfrac{\delta R^2_0}{L_0 \sigma_L}\right)}.
\ee
Next, using $N \gg 1$ instead of $\xi$, we arrive at the following estimate of the boundary between extended and localized regimes:
\be
  \frac{\delta R^2_0}{L_0 \sigma_L} \sim \ln{N},
\ee
which \textit{a posteriori} justifies our assumption $\delta R^2_0 \gg L_0 \sigma_L$. This boundary has been indeed observed in our numerics, see Fig.~\ref{fig_1D_phase_diag}, where the extended regime is visible in a certain domain of parameters in the vicinity of the vertical axis. The system size here is $N=120$.




\begin{figure}[h]
  \includegraphics[width=3.5cm]{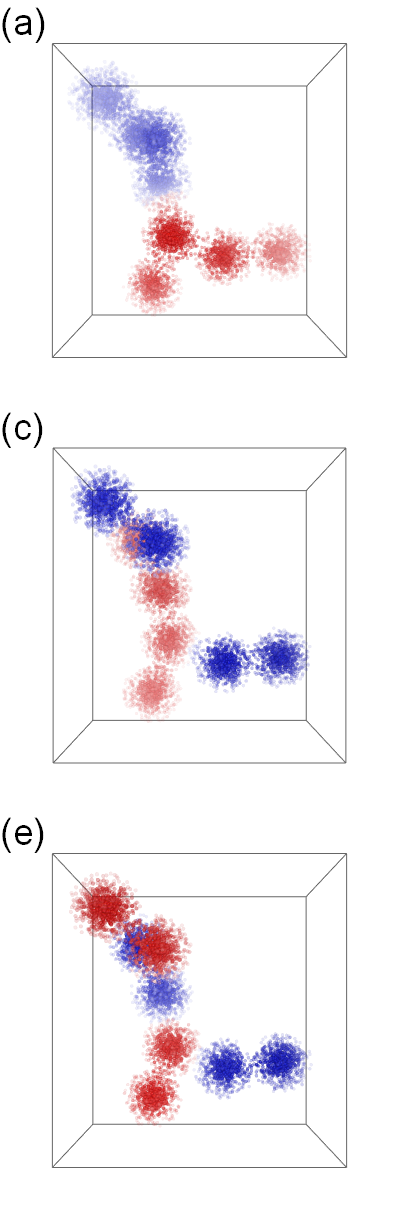}\includegraphics[width=3.5cm]{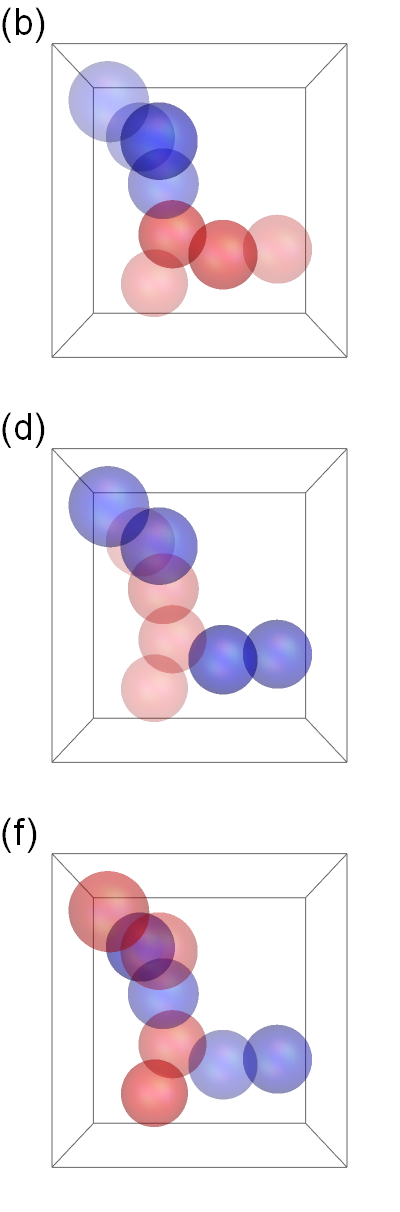}\\
  \vspace{8pt}
  \includegraphics[width=7.6cm]{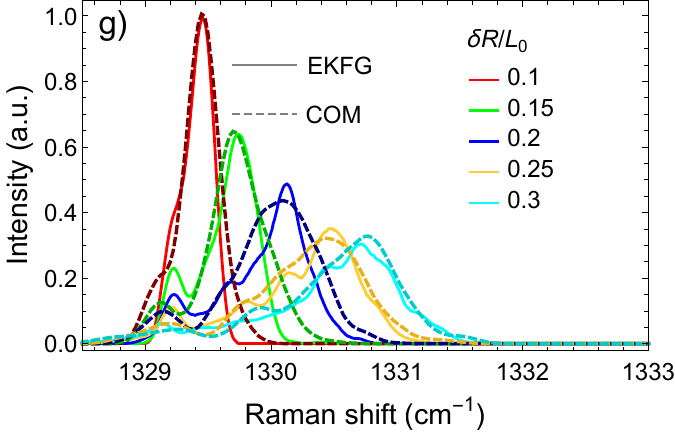}
   \caption{\label{fig_aggl8p_COMvsEKFG} The second, the third, and the fourth eigenfunctions of the 8-particle agglomerate obtained within EKFG (a,c,e) and COM (b,d,f). Panel (g) shows the Raman spectra of the ensembles of 8-nanoparticle agglomerates calculated within EKFG (dashed curves) and COM (solid curves) approaches with various values of the penetration depth $\delta R$. The mean size of particles is $L_0=3.1$~nm. }
\end{figure}

\subsection{Tight agglomerates}

\label{subsec_agglomerates}

We proceed with another type of physical system -- the tight agglomerate of nanoparticles. Even for identical particles, it can be considered as a disordered array due to its spatial structure. Importantly, agglomerates can be characterized by their fractal dimension~\cite{meakin1984effects,brasil1999recipe,brasil2000evaluation}, which is governed by the underlying formation mechanism. The fractal dimension indicated below is defined via the slope in the dependence of the gyration radius versus the number of particles belonging to the agglomerate.

In the present study, the agglomeration process is modeled according to the procedures described in Ref.~\cite{brasil2001numerical}. In the case of the cluster-cluster aggregation model, both collision and adhesion take place for agglomerates of similar size. It results in a doubling of the structure size on each step. For particle-cluster aggregation, the single particles hit the growing agglomerate. During the cluster-cluster process, the resulting agglomerates are formed more sparsely (the corresponding fractal dimension $D_f \approx 1.8$) than in the case of the particle-cluster process where $D_f\approx 2.7$. The fractal dimension of detonation nanodiamond agglomerates obtained from the neutron scattering experiments and optical measurements has a value close to 2.5~\cite{avdeev2009aggregate,tomchuk2015structural,koniakhin2020evidence}, which indicates that the particle-cluster model is more relevant to that kind of systems.

Importantly, when utilizing the EKFG approach, we observe a peculiarity. Even considering the particles of the same size and penetration depth of nearest neighbors, we found that there is a variance in the particle overlap volumes, which naturally occurs due to the finite size of elements upon creating the mesh for the solution of \mbox{$\Delta \psi + q^2 \psi = 0$} eigenvalue problem in Wolfram Mathematica package~\cite{Mathematica}. To take it into account in COM calculations, we introduce a random coefficient for coupling terms with the unit mean value and the variance equal to $0.5$.

In panels (a)-(f) of Fig. \ref{fig_aggl8p_COMvsEKFG} we compare the second, the third, and the fourth eigenfunctions obtained using EKFG and COM approaches for agglomerates containing 8 particles of the same size $L_0 = 3.1$~nm (the first eigenfunctions which are the ``boring'' constant sign solutions are not shown here). One can see an excellent agreement in the wave functions' spatial structure between the employed methods. For EKFG results, the random points belonging to the agglomerate are taken to plot the square of the wave function amplitude that is depicted proportionally to the intensity of color. For the COM approach, the squared wave function amplitude is proportional to the color intensity as well. Red and blue colors indicate different relative signs of the wave functions. Panel (g) shows the Raman spectra of the ensembles of 300 agglomerates for various relative overlapping parameters $\delta R/L_0$ calculated using COM and EKFG approaches. One can see a good agreement between the methods. Thus, we can safely use COM for studying the properties of larger agglomerates, hardly suitable for EKFG calculations.

\begin{figure}
  \includegraphics[width=7.6cm]{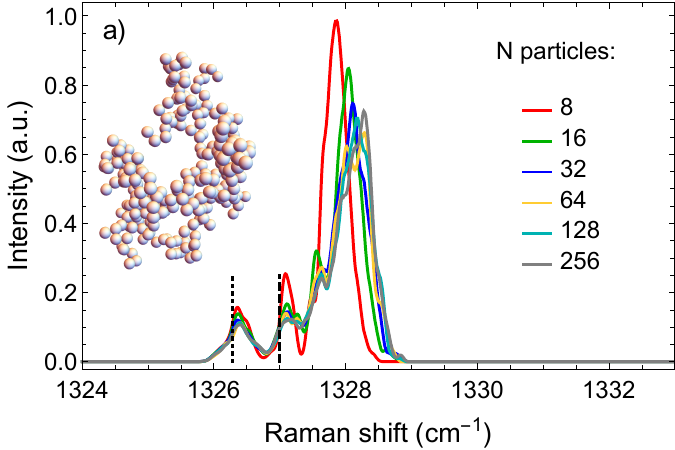}\\
   \includegraphics[width=7.6cm]{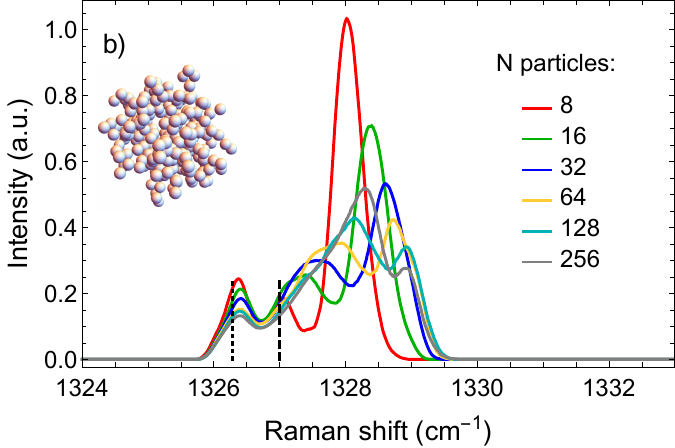}
   \caption{\label{fig_COM_allglomerate_size}Raman spectra of various-sized tight agglomerates. The size of diamond nanoparticles is 2.4~nm. Various colors correspond to agglomerates of various sizes. Isolated particle and dimer peak positions are indicated by dotted and dashed lines, respectively. Panel (a) shows the results for the cluster-cluster aggregation procedure and panel (b) shows the results for the cluster-particle aggregation. The insets demonstrate the typical structure of the agglomerate consisting of 256 particles.}
\end{figure}

In Fig. \ref{fig_COM_allglomerate_size} we show the evolution of the Raman spectra of agglomerates constructed using cluster-cluster and cluster-particle mechanisms upon the growth in the number of particles. In these calculations, we employ COM and use identical particles with $L_0 = 2.4$~nm and $\delta R/L_0 = 0.2$. One can see that in both cases the peak shifts towards higher frequencies when the particle number increases. Notice that for the cluster-particle mechanism, the peak becomes lower and broader than in the cluster-cluster case. Importantly, in both cases, the main peak corresponds to the collective modes of the agglomerate. Contributions from modes localized on a single particle or two neighboring particles are at lower frequencies and are much weaker.

\begin{figure}
  \includegraphics[width=6cm]{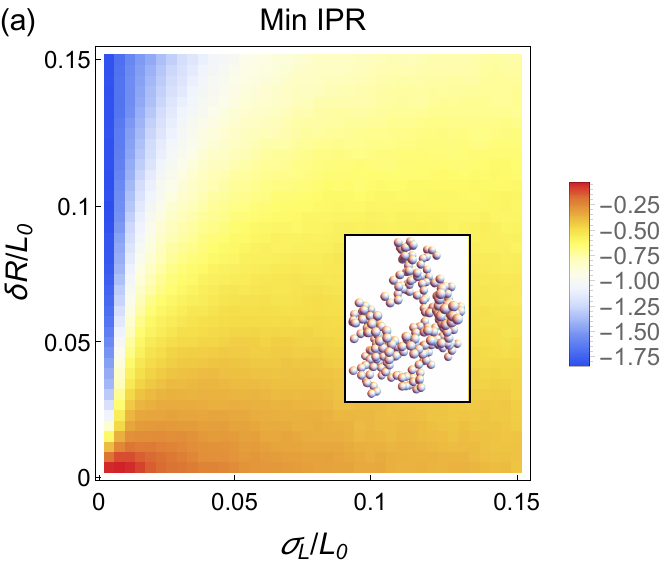}
  \includegraphics[width=6cm]{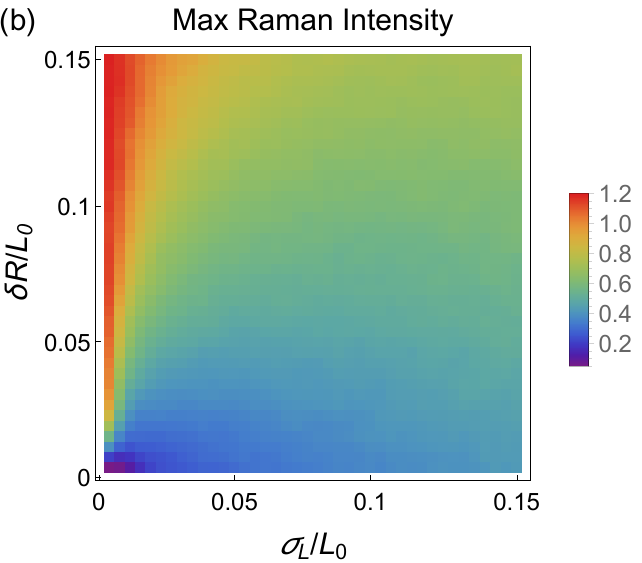}\\
   \includegraphics[width=6cm]{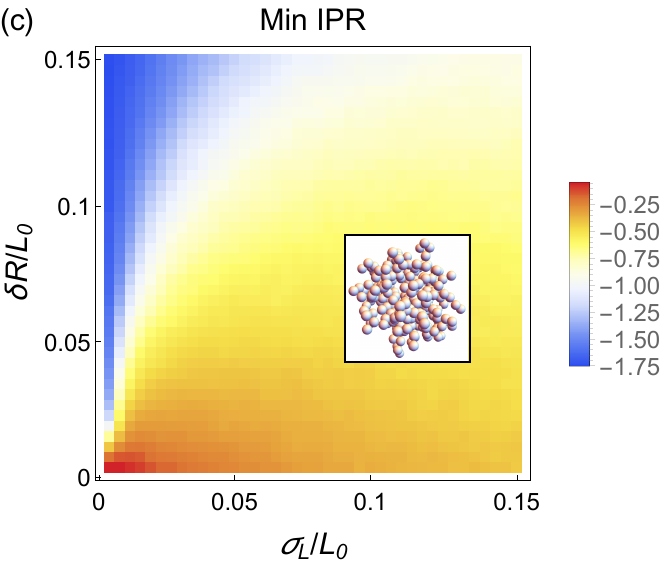}
   \includegraphics[width=6cm]{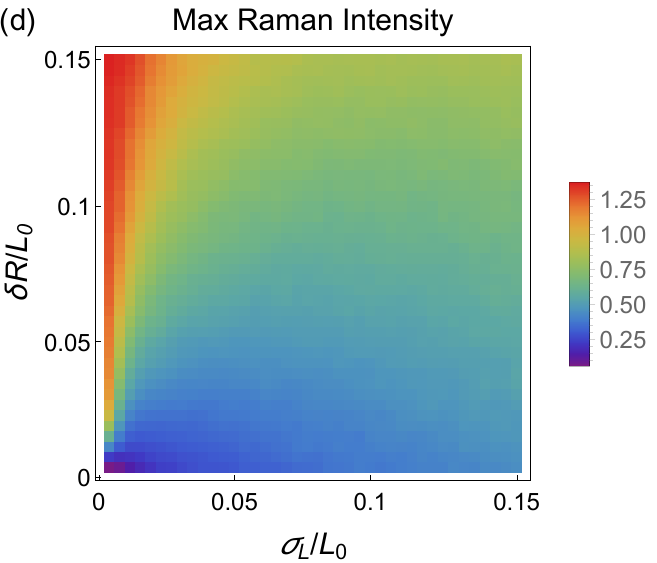}\\
   \caption{\label{fig_IPR_aggl} IPR and the maximal Raman intensity for cluster-cluster [(a) and (b)] and cluster-particle [(c) and (d)] agglomerates consisting of 256 particles.}
\end{figure}

Now we turn to the characterization of eigenmodes obtained for two types of agglomerates with disorder in particle sizes using COM. Our findings are presented in Fig.~\ref{fig_IPR_aggl}. One can see that the minimal IPR  and the maximal Raman intensity of a single mode behave in a qualitatively similar way to 3D and 2D porous media cases (cf. Figs.~\ref{fig_3D_phase_diag} and~~\ref{fig_2D_phase_diag}). When the particle size variance grows, the eigenmodes become essentially localized, and their Raman intensities drop.

\begin{figure}
    \includegraphics[width=8cm]{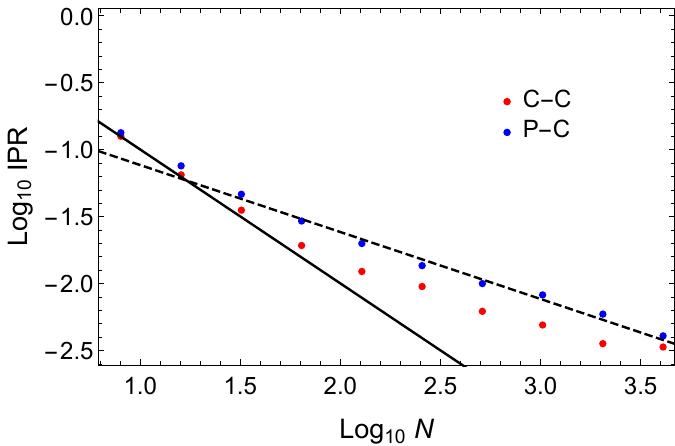}\\
  \caption{Minimal IPR of a single mode versus the number of particles in agglomerate $N$ for cluster-cluster and particle-cluster agglomerates. The lines give a guide an the eye for $y\propto1/x$ (solid line) and $y\propto1/\sqrt{x}$ (dashed line).}
  \label{fig_IPRvsN}
\end{figure}

To further check the properties of agglomerate modes with minimal IPR in the absence of the disorder, we study IPR evolution with the number of particles $N$. We observe two different scenarios, see Fig.~\ref{fig_IPRvsN}. In the case of the cluster-cluster mechanism, IPR tends to saturate at large $N$ thus the modes are localized, whereas for the cluster-particle mechanism, IPR corresponds to extended fractal states with $\alpha \approx 0.5$ [see text after Eq.~\eqref{IPR}; this parameter should not be confused with the fractal dimension of the agglomerate $D_f$]. We believe, that this behavior can be understood by comparison of the agglomerates' $D_f$ parameters with the critical dimension for Anderson transition $D=2$ which divides regimes of localized and extended states in weakly disordered systems~\cite{abrahams2010}. However, we note that the problem of (de)localization of elementary excitations in random fractal structures of large size $N$ deserves a separate comprehensive study (for regular fractals see, e.g., Refs.~\cite{rammal1983nature,wang1995localization,pal2012}).

\begin{figure}
    \includegraphics[width=8cm]{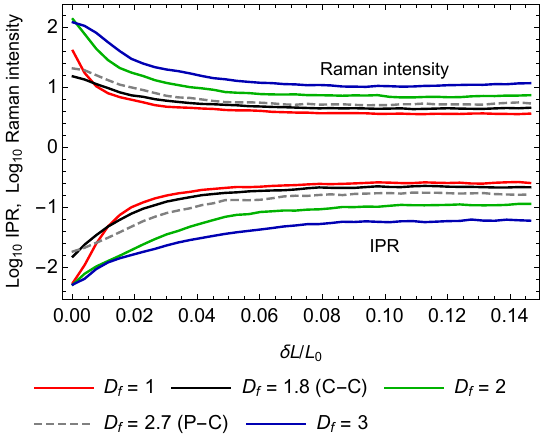}\\
  \caption{Maximal Raman intensity and minimal IPR of a single mode versus particle size variance for various types of nanoparticle arrays. 1D chain, regular 2D (256 particles) and 3D (216 particles) arrays, cluster-cluster agglomerates, and particle-cluster agglomerates (256 particles) characterized by various fractal dimensions are considered.}
  \label{fig_ramanIPRvsdL}
\end{figure}

Finally, we summarize the behavior of minimal IPR and maximal Raman intensity of a single mode for various fractal dimensions $D_f$ of the finite systems with $N \sim 250$ particles as functions of the particle size variance in Fig.~\ref{fig_ramanIPRvsdL}. One can see that at $\sigma_L=0$ the curves start from purely extended states in ordered systems (integer $D_f$) and in the agglomerates with non-integer $D_f$. However, we find that in the former case $\alpha=1$ and in the latter case $\alpha < 1$.  Then, upon the $\sigma_L$ growth the most dramatic effect is visible for the 1D chain, whereas for other systems the effects of Raman intensity of a single mode decrease and the IPR increase are much more gradual.

\section{Discussion and conclusions}
\label{sec_concl}

This paper is devoted to the description of the collective optical phonon modes in several physically or chemically implementable structured materials. These materials are assumed to consist of smaller objects (nanoparticles) mutually connected and arranged into various types of arrays (ordered and disordered nanosolids of certain dimensionality, porous media,  fractal agglomerates, etc.) by means of ``weak inter-particle links''. The latter provides the mechanism of propagation throughout an array for otherwise intrinsic (confined) modes due to the hybridization of phonons with close energies that belong to different contacting particles. In order to optimize the calculations the simple coupled-oscillator model is used instead of more involved and costly DMM or EKFG approaches, the COM is shown to be working quite well for the problem at hand. The above theory allows us to judge confidently the localization/delocalization of optical modes on scales exceeding the particle size $L$. The modification of Raman spectra due to the above phenomena is also discussed with reasonable accuracy.

To be more specific, the regular 3D arrays of nanoparticles  (ordered and disordered) are proposed as a proper model of nanosolids and porous media. For
disordered arrays, the inverse participation ratio is investigated as a function of the overlap parameter (measuring the inter-mode hybridization) and
the particle size variance (responsible for the disorder). The detailed analysis allows us to obtain analytically the scale of localization/delocalization transition and to confirm this result by numerical calculations. This 3D analysis is accompanied by the model calculations of 2D and 1D arrays. The similarity of the problem at hand with the Anderson model with on-site disorder and with the percolation scenario is emphasized. The Raman intensity is evaluated as a function of the above parameters, and the line-broadening enhancement in the localized regime is attributed.
The tight agglomerates of nanoparticles are presented in the forms of two (cluster-cluster and cluster-particle) models of aggregation. The influence of the fractal structure of agglomerates on the localization properties of optical phonons is investigated via the IPR study, and the Raman shift and line broadening in such systems are depicted as the functions of disorder and coupling parameters.

The present paper paves the way toward including many-particle collective effects in previously developed efficient methods for nanoparticle ensemble Raman spectra calculations. These previously neglected effects can significantly modify the results and thus the interpretation of the corresponding experimental data.


\begin{acknowledgments}

We are grateful to S. Flach and A. Andreanov for their attention to our research and to B. Altshuler for stimulating discussions. S. V. K. acknowledges financial support from the Institute for Basic Science (IBS, Korea) Young Scientist Fellowship (IBS-R024-Y3). O. I. U. acknowledges financial support from the Institute for Basic Science in Korea (Project No. IBS-R024-D1) and from the Foundation for the Advancement of Theoretical Physics and Mathematics ``Basis''.

\end{acknowledgments}

\bibliography{main}

\begin{thebibliography}{10}
\providecommand{\url}[1]{#1}
\csname url@samestyle\endcsname
\providecommand{\newblock}{\relax}
\providecommand{\bibinfo}[2]{#2}
\providecommand{\BIBentrySTDinterwordspacing}{\spaceskip=0pt\relax}
\providecommand{\BIBentryALTinterwordstretchfactor}{4}
\providecommand{\BIBentryALTinterwordspacing}{\spaceskip=\fontdimen2\font plus
\BIBentryALTinterwordstretchfactor\fontdimen3\font minus \fontdimen4\font\relax}
\providecommand{\BIBforeignlanguage}[2]{{%
\expandafter\ifx\csname l@#1\endcsname\relax
\typeout{** WARNING: IEEEtran.bst: No hyphenation pattern has been}%
\typeout{** loaded for the language `#1'. Using the pattern for}%
\typeout{** the default language instead.}%
\else
\language=\csname l@#1\endcsname
\fi
#2}}
\providecommand{\BIBdecl}{\relax}
\BIBdecl

\bibitem{khan2019nanoparticles}
I.~Khan, K.~Saeed, and I.~Khan, ``Nanoparticles: Properties, applications and toxicities,'' \emph{Arabian journal of chemistry}, vol.~12, no.~7, pp. 908--931, 2019.

\bibitem{mochalin2020properties}
V.~Mochalin, O.~Shenderova, D.~Ho, and Y.~Gogotsi, ``The properties and applications of nanodiamonds,'' \emph{Nano-enabled medical applications}, pp. 313--350, 2020.

\bibitem{Mitchell2021}
\BIBentryALTinterwordspacing
M.~J. Mitchell, M.~M. Billingsley, R.~M. Haley, M.~E. Wechsler, N.~A. Peppas, and R.~Langer, ``Engineering precision nanoparticles for drug delivery,'' \emph{Nature Reviews Drug Discovery}, vol.~20, no.~2, pp. 101--124, Feb 2021. [Online]. Available: \url{https://doi.org/10.1038/s41573-020-0090-8}
\BIBentrySTDinterwordspacing

\bibitem{astruc2020introduction}
D.~Astruc, ``Introduction: nanoparticles in catalysis,'' pp. 461--463, 2020.

\bibitem{chen2021atomically}
X.~Chen, Z.~Jia, F.~Huang, J.~Diao, and H.~Liu, ``Atomically dispersed metal catalysts on nanodiamond and its derivatives: synthesis and catalytic application,'' \emph{Chemical Communications}, vol.~57, no.~88, pp. 11\,591--11\,603, 2021.

\bibitem{terna2021future}
A.~D. Terna, E.~E. Elemike, J.~I. Mbonu, O.~E. Osafile, and R.~O. Ezeani, ``The future of semiconductors nanoparticles: Synthesis, properties and applications,'' \emph{Materials Science and Engineering: B}, vol. 272, p. 115363, 2021.

\bibitem{chernikov2023tunable}
A.~S. Chernikov, G.~I. Tselikov, M.~Y. Gubin, A.~V. Shesterikov, K.~S. Khorkov, A.~V. Syuy, G.~A. Ermolaev, I.~S. Kazantsev, R.~I. Romanov, A.~M. Markeev \emph{et~al.}, ``Tunable optical properties of transition metal dichalcogenide nanoparticles synthesized by femtosecond laser ablation and fragmentation,'' \emph{Journal of Materials Chemistry C}, vol.~11, no.~10, pp. 3493--3503, 2023.

\bibitem{nunn2023optical}
N.~Nunn, S.~Milikisiyants, M.~D. Torelli, R.~Monge, T.~Delord, A.~I. Shames, C.~A. Meriles, A.~Ajoy, A.~I. Smirnov, and O.~A. Shenderova, ``Optical and electronic spin properties of fluorescent micro-and nanodiamonds upon prolonged ultrahigh-temperature annealing,'' \emph{Journal of Vacuum Science \& Technology B}, vol.~41, no.~4, 2023.

\bibitem{bao2010biosynthesis}
H.~Bao, N.~Hao, Y.~Yang, and D.~Zhao, ``Biosynthesis of biocompatible cadmium telluride quantum dots using yeast cells,'' \emph{Nano Research}, vol.~3, pp. 481--489, 2010.

\bibitem{fusco2020impact}
L.~Fusco, E.~Avitabile, V.~Armuzza, M.~Orecchioni, A.~Istif, D.~Bedognetti, T.~Da~Ros, and L.~G. Delogu, ``Impact of the surface functionalization on nanodiamond biocompatibility: A comprehensive view on human blood immune cells,'' \emph{Carbon}, vol. 160, pp. 390--404, 2020.

\bibitem{leung2023versatile}
H.~M. Leung, H.~C. Chu, Z.-W. Mao, and P.~K. Lo, ``Versatile nanodiamond-based tools for therapeutics and bioimaging,'' \emph{Chemical Communications}, vol.~59, no.~15, pp. 2039--2055, 2023.

\bibitem{das2023polymer}
S.~K. Das, L.~Pradhan, B.~K. Jena, and S.~Basu, ``Polymer derived honeycomb-like carbon nanostructures for high capacitive supercapacitor application,'' \emph{Carbon}, vol. 201, pp. 49--59, 2023.

\bibitem{nyholm2023functionalized}
N.~Nyholm and N.~Espallargas, ``Functionalized carbon nanostructures as lubricant additives--a review,'' \emph{Carbon}, vol. 201, pp. 1200--1228, 2023.

\bibitem{mochalin2012properties}
V.~N. Mochalin, O.~Shenderova, D.~Ho, and Y.~Gogotsi, ``The properties and applications of nanodiamonds,'' \emph{Nature nanotechnology}, vol.~7, no.~1, pp. 11--23, 2012.

\bibitem{wood2022long}
B.~Wood, G.~Stimpson, J.~March, Y.~Lekhai, C.~Stephen, B.~Green, A.~Frangeskou, L.~Gin{\'e}s, S.~Mandal, O.~Williams \emph{et~al.}, ``Long spin coherence times of nitrogen vacancy centers in milled nanodiamonds,'' \emph{Physical Review B}, vol. 105, no.~20, p. 205401, 2022.

\bibitem{panfil2019electronic}
Y.~E. Panfil, D.~Shamalia, J.~Cui, S.~Koley, and U.~Banin, ``Electronic coupling in colloidal quantum dot molecules; the case of cdse/cds core/shell homodimers,'' \emph{The Journal of chemical physics}, vol. 151, no.~22, p. 224501, 2019.

\bibitem{cui2019colloidal}
J.~Cui, Y.~E. Panfil, S.~Koley, D.~Shamalia, N.~Waiskopf, S.~Remennik, I.~Popov, M.~Oded, and U.~Banin, ``Colloidal quantum dot molecules manifesting quantum coupling at room temperature,'' \emph{Nature communications}, vol.~10, no.~1, pp. 1--10, 2019.

\bibitem{koley2021coupled}
S.~Koley, J.~Cui, Y.~E. Panfil, and U.~Banin, ``Coupled colloidal quantum dot molecules,'' \emph{Accounts of Chemical Research}, vol.~54, no.~5, pp. 1178--1188, 2021.

\bibitem{cui2021neck}
J.~Cui, S.~Koley, Y.~E. Panfil, A.~Levi, Y.~Ossia, N.~Waiskopf, S.~Remennik, M.~Oded, and U.~Banin, ``Neck barrier engineering in quantum dot dimer molecules via intraparticle ripening,'' \emph{Journal of the American Chemical Society}, vol. 143, no.~47, pp. 19\,816--19\,823, 2021.

\bibitem{cui2021semiconductor}
J.~Cui, S.~Koley, Y.~E. Panfil, A.~Levi, N.~Waiskopf, S.~Remennik, M.~Oded, and U.~Banin, ``Semiconductor bow-tie nanoantenna from coupled colloidal quantum dot molecules,'' \emph{Angewandte Chemie International Edition}, 2021.

\bibitem{bozyigit2014electrical}
D.~Bozyigit and V.~Wood, ``Electrical characterization of nanocrystal solids,'' \emph{Journal of Materials Chemistry C}, vol.~2, no.~17, pp. 3172--3184, 2014.

\bibitem{zhao2021enhanced}
Q.~Zhao, G.~Gouget, J.~Guo, S.~Yang, T.~Zhao, D.~B. Straus, C.~Qian, N.~Oh, H.~Wang, C.~B. Murray \emph{et~al.}, ``Enhanced carrier transport in strongly coupled, epitaxially fused cdse nanocrystal solids,'' \emph{Nano Letters}, vol.~21, no.~7, pp. 3318--3324, 2021.

\bibitem{jang2015solution}
J.~Jang, D.~S. Dolzhnikov, W.~Liu, S.~Nam, M.~Shim, and D.~V. Talapin, ``Solution-processed transistors using colloidal nanocrystals with composition-matched molecular “solders”: approaching single crystal mobility,'' \emph{Nano letters}, vol.~15, no.~10, pp. 6309--6317, 2015.

\bibitem{evers2015high}
W.~H. Evers, J.~M. Schins, M.~Aerts, A.~Kulkarni, P.~Capiod, M.~Berthe, B.~Grandidier, C.~Delerue, H.~S. Van Der~Zant, C.~Van~Overbeek \emph{et~al.}, ``High charge mobility in two-dimensional percolative networks of pbse quantum dots connected by atomic bonds,'' \emph{Nature communications}, vol.~6, no.~1, p. 8195, 2015.

\bibitem{whitham2016charge}
K.~Whitham, J.~Yang, B.~H. Savitzky, L.~F. Kourkoutis, F.~Wise, and T.~Hanrath, ``Charge transport and localization in atomically coherent quantum dot solids,'' \emph{Nature materials}, vol.~15, no.~5, pp. 557--563, 2016.

\bibitem{chen2016metal}
T.~Chen, K.~Reich, N.~J. Kramer, H.~Fu, U.~R. Kortshagen, and B.~Shklovskii, ``Metal--insulator transition in films of doped semiconductor nanocrystals,'' \emph{Nature materials}, vol.~15, no.~3, pp. 299--303, 2016.

\bibitem{reich2016exciton}
K.~V. Reich and B.~I. Shklovskii, ``Exciton transfer in array of epitaxially connected nanocrystals,'' \emph{ACS nano}, vol.~10, no.~11, pp. 10\,267--10\,274, 2016.

\bibitem{yang1994study}
M.~Yang, D.~Huang, P.~Hao, F.~Zhang, X.~Hou, and X.~Wang, ``Study of the raman peak shift and the linewidth of light-emitting porous silicon,'' \emph{Journal of applied physics}, vol.~75, no.~1, pp. 651--653, 1994.

\bibitem{alfaro2008theory}
P.~Alfaro-Calder{\'o}n, M.~Cruz-Irisson, and C.~Wang-Chen, ``Theory of raman scattering by phonons in germanium nanostructures,'' \emph{Nanoscale Research Letters}, vol.~3, no.~2, pp. 55--59, 2008.

\bibitem{alfaro2011raman}
P.~Alfaro, R.~Cisneros, M.~Bizarro, M.~Cruz-Irisson, and C.~Wang, ``Raman scattering by confined optical phonons in si and ge nanostructures,'' \emph{Nanoscale}, vol.~3, no.~3, pp. 1246--1251, 2011.

\bibitem{valtchev2013porous}
V.~Valtchev and L.~Tosheva, ``Porous nanosized particles: preparation, properties, and applications,'' \emph{Chemical Reviews}, vol. 113, no.~8, pp. 6734--6760, 2013.

\bibitem{kosovic2014phonon}
M.~Kosovi{\'c}, O.~Gamulin, M.~Balarin, M.~Ivanda, V.~Derek, D.~Risti{\'c}, M.~Marciu{\v{s}}, and M.~Risti{\'c}, ``Phonon confinement effects in raman spectra of porous silicon at non-resonant excitation condition,'' \emph{Journal of Raman Spectroscopy}, vol.~45, no.~6, pp. 470--475, 2014.

\bibitem{raty2003ultradispersity}
J.-Y. Raty and G.~Galli, ``Ultradispersity of diamond at the nanoscale,'' \emph{Nature materials}, vol.~2, no.~12, pp. 792--795, 2003.

\bibitem{ozerin2008x}
A.~Ozerin, T.~Kurkin, L.~Ozerina, and V.~Y. Dolmatov, ``X-ray diffraction study of the structure of detonation nanodiamonds,'' \emph{Crystallography Reports}, vol.~53, no.~1, pp. 60--67, 2008.

\bibitem{koniakhin2015molecular}
S.~Koniakhin, I.~Eliseev, I.~Terterov, A.~Shvidchenko, E.~Eidelman, and M.~Dubina, ``Molecular dynamics-based refinement of nanodiamond size measurements obtained with dynamic light scattering,'' \emph{Microfluidics and Nanofluidics}, vol.~18, no.~5, pp. 1189--1194, 2015.

\bibitem{koniakhin2018ultracentrifugation}
S.~Koniakhin, N.~Besedina, D.~Kirilenko, A.~Shvidchenko, and E.~Eidelman, ``Ultracentrifugation for ultrafine nanodiamond fractionation,'' \emph{Superlattices and Microstructures}, vol. 113, pp. 204--212, 2018.

\bibitem{trofimuk2018effective}
A.~D. Trofimuk, D.~V. Muravijova, D.~A. Kirilenko, and A.~V. Shvidchenko, ``Effective method for obtaining the hydrosols of detonation nanodiamond with particle size< 4 nm,'' \emph{Materials}, vol.~11, no.~8, p. 1285, 2018.

\bibitem{baidakova1999ultradisperse}
M.~Baidakova, V.~Siklitsky, and A.~Y. Vul, ``Ultradisperse-diamond nanoclusters. fractal structure and diamond--graphite phase transition,'' \emph{Chaos, Solitons \& Fractals}, vol.~10, no.~12, pp. 2153--2163, 1999.

\bibitem{williams2007enhanced}
O.~A. Williams, O.~Douh{\'e}ret, M.~Daenen, K.~Haenen, E.~{\=O}sawa, and M.~Takahashi, ``Enhanced diamond nucleation on monodispersed nanocrystalline diamond,'' \emph{Chemical Physics Letters}, vol. 445, no. 4-6, pp. 255--258, 2007.

\bibitem{osawa2008monodisperse}
E.~{\=O}sawa, ``Monodisperse single nanodiamond particulates,'' \emph{Pure and Applied Chemistry}, vol.~80, no.~7, pp. 1365--1379, 2008.

\bibitem{avdeev2009aggregate}
M.~Avdeev, N.~Rozhkova, V.~Aksenov, V.~Garamus, R.~Willumeit, and E.~Osawa, ``Aggregate structure in concentrated liquid dispersions of ultrananocrystalline diamond by small-angle neutron scattering,'' \emph{The Journal of Physical Chemistry C}, vol. 113, no.~22, pp. 9473--9479, 2009.

\bibitem{tomchuk2015structural}
O.~V. Tomchuk, D.~S. Volkov, L.~A. Bulavin, A.~V. Rogachev, M.~A. Proskurnin, M.~V. Korobov, and M.~V. Avdeev, ``Structural characteristics of aqueous dispersions of detonation nanodiamond and their aggregate fractions as revealed by small-angle neutron scattering,'' \emph{The Journal of Physical Chemistry C}, vol. 119, no.~1, pp. 794--802, 2015.

\bibitem{dideikin2017rehybridization}
A.~Dideikin, A.~Aleksenskii, M.~Baidakova, P.~Brunkov, M.~Brzhezinskaya, V.~Y. Davydov, V.~Levitskii, S.~Kidalov, Y.~A. Kukushkina, D.~Kirilenko \emph{et~al.}, ``Rehybridization of carbon on facets of detonation diamond nanocrystals and forming hydrosols of individual particles,'' \emph{Carbon}, vol. 122, pp. 737--745, 2017.

\bibitem{koniakhin2020evidence}
S.~Koniakhin, M.~Rabchinskii, N.~Besedina, L.~Sharonova, A.~Shvidchenko, and E.~Eidelman, ``Evidence of absorption dominating over scattering in light attenuation by nanodiamonds,'' \emph{Physical Review Research}, vol.~2, no.~1, p. 013316, 2020.

\bibitem{korepanov2017carbon}
V.~I. Korepanov, H.-o. Hamaguchi, E.~Osawa, V.~Ermolenkov, I.~K. Lednev, B.~J. Etzold, O.~Levinson, B.~Zousman, C.~P. Epperla, and H.-C. Chang, ``Carbon structure in nanodiamonds elucidated from raman spectroscopy,'' \emph{Carbon}, vol. 121, pp. 322--329, 2017.

\bibitem{vinogradov2018structure}
A.~Y. Vinogradov, S.~Grudinkin, N.~Besedina, S.~Koniakhin, M.~Rabchinskii, E.~Eidelman, and V.~Golubev, ``Structure and properties of thin graphite-like films produced by magnetron-assisted sputtering,'' \emph{Semiconductors}, vol.~52, no.~7, pp. 914--920, 2018.

\bibitem{korepanov2020localized}
V.~I. Korepanov, ``Localized phonons in raman spectra of nanoparticles and disordered media,'' \emph{Journal of Raman Spectroscopy}, vol.~51, no.~6, pp. 881--891, 2020.

\bibitem{qiu2022two}
J.~Qiu, T.~H. Nguyen, S.~Kim, Y.~J. Lee, M.-T. Song, W.-J. Huang, X.-B. Chen, T.~M.~H. Nguyen, and I.-S. Yang, ``Two-dimensional correlation spectroscopy analysis of raman spectra of nio nanoparticles,'' \emph{Spectrochimica Acta Part A: Molecular and Biomolecular Spectroscopy}, vol. 280, p. 121498, 2022.

\bibitem{zucker1987resonant}
J.~Zucker, A.~Pinczuk, D.~Chemla, and A.~Gossard, ``Resonant raman study of low-temperature exciton localization in gaas quantum wells,'' \emph{Physical Review B}, vol.~35, no.~6, p. 2892, 1987.

\bibitem{brewster2009exciton}
M.~Brewster, O.~Schimek, S.~Reich, and S.~Grade{\v{c}}ak, ``Exciton-phonon coupling in individual gaas nanowires studied using resonant raman spectroscopy,'' \emph{Physical Review B}, vol.~80, no.~20, p. 201314, 2009.

\bibitem{ferrari2004raman}
A.~C. Ferrari and J.~Robertson, ``Raman spectroscopy of amorphous, nanostructured, diamond--like carbon, and nanodiamond,'' \emph{Philosophical Transactions of the Royal Society of London. Series A: Mathematical, Physical and Engineering Sciences}, vol. 362, no. 1824, pp. 2477--2512, 2004.

\bibitem{osswald2009phonon}
S.~Osswald, V.~Mochalin, M.~Havel, G.~Yushin, and Y.~Gogotsi, ``Phonon confinement effects in the raman spectrum of nanodiamond,'' \emph{Physical Review B}, vol.~80, no.~7, p. 075419, 2009.

\bibitem{kumar2012raman}
C.~S. Kumar, \emph{Raman spectroscopy for nanomaterials characterization}.\hskip 1em plus 0.5em minus 0.4em\relax Springer Science \& Business Media, 2012.

\bibitem{lee2019raman}
T.~Lee, J.~S. Kim, S.~J. Lee, and H.~Rho, ``Raman study of inas/gaas quantum dot solar cells,'' \emph{Current Applied Physics}, vol.~19, no.~10, pp. 1132--1135, 2019.

\bibitem{ourBench}
\BIBentryALTinterwordspacing
A.~G. Yashenkin, O.~I. Utesov, and S.~V. Koniakhin, ``Bench tests for microscopic theory of raman scattering in powders of disordered nonpolar crystals: Nanodiamonds and beyond,'' \emph{Journal of Raman Spectroscopy}, vol.~52, pp. 1847--1859, 2021. [Online]. Available: \url{https://analyticalsciencejournals.onlinelibrary.wiley.com/doi/abs/10.1002/jrs.6242}
\BIBentrySTDinterwordspacing

\bibitem{kashyap2021comparative}
V.~Kashyap, C.~Kumar, N.~Chaudhary, N.~Goyal, and K.~Saxena, ``Comparative study of quantum confinements effect present in silicon nanowires using absorption and raman spectroscopy,'' \emph{Optical Materials}, vol. 121, p. 111538, 2021.

\bibitem{koniakhin2023coupled}
S.~Koniakhin, O.~Utesov, and A.~Yashenkin, ``Coupled-oscillator model for hybridized optical phonon modes in contacting nanosize particles and quantum dot molecules,'' \emph{Physical Review Research}, vol.~5, no.~1, p. 013153, 2023.

\bibitem{ourDMM}
\BIBentryALTinterwordspacing
S.~V. Koniakhin, O.~I. Utesov, I.~N. Terterov, A.~V. Siklitskaya, A.~G. Yashenkin, and D.~Solnyshkov, ``Raman spectra of crystalline nanoparticles: Replacement for the phonon confinement model,'' \emph{The Journal of Physical Chemistry C}, vol. 122, no.~33, pp. 19\,219--19\,229, 2018. [Online]. Available: \url{https://doi.org/10.1021/acs.jpcc.8b05415}
\BIBentrySTDinterwordspacing

\bibitem{ourEKFG}
\BIBentryALTinterwordspacing
O.~I. Utesov, A.~G. Yashenkin, and S.~V. Koniakhin, ``Raman spectra of nonpolar crystalline nanoparticles: Elasticity theory-like approach for optical phonons,'' \emph{The Journal of Physical Chemistry C}, vol. 122, no.~39, pp. 22\,738--22\,749, 2018. [Online]. Available: \url{https://doi.org/10.1021/acs.jpcc.8b07061}
\BIBentrySTDinterwordspacing

\bibitem{anderson1958absence}
P.~W. Anderson, ``Absence of diffusion in certain random lattices,'' \emph{Physical review}, vol. 109, no.~5, p. 1492, 1958.

\bibitem{abrahams2010}
\BIBentryALTinterwordspacing
E.~Abrahams, \emph{50 Years of Anderson Localization}, ser. International journal of modern physics: Condensed matter physics, statistical physics, applied physics.\hskip 1em plus 0.5em minus 0.4em\relax World Scientific, 2010. [Online]. Available: \url{https://books.google.co.kr/books?id=6O5pDQAAQBAJ}
\BIBentrySTDinterwordspacing

\bibitem{richter1981one}
H.~Richter, Z.~Wang, and L.~Ley, ``The one phonon raman spectrum in microcrystalline silicon,'' \emph{Solid State Communications}, vol.~39, no.~5, pp. 625--629, 1981.

\bibitem{campbell1986effects}
I.~Campbell and P.~M. Fauchet, ``The effects of microcrystal size and shape on the one phonon raman spectra of crystalline semiconductors,'' \emph{Solid State Communications}, vol.~58, no.~10, pp. 739--741, 1986.

\bibitem{jansen2023nanocrystal}
M.~Jansen, W.~A. Tisdale, and V.~Wood, ``Nanocrystal phononics,'' \emph{Nature Materials}, vol.~22, no.~2, pp. 161--169, 2023.

\bibitem{yazdani2019nanocrystal}
N.~Yazdani, M.~Jansen, D.~Bozyigit, W.~M. Lin, S.~Volk, O.~Yarema, M.~Yarema, F.~Juranyi, S.~D. Huber, and V.~Wood, ``Nanocrystal superlattices as phonon-engineered solids and acoustic metamaterials,'' \emph{Nature Communications}, vol.~10, no.~1, p. 4236, 2019.

\bibitem{jansen2019phonon}
M.~Jansen, N.~Yazdani, and V.~Wood, ``Phonon-engineered solids constructed from nanocrystals,'' \emph{APL Materials}, vol.~7, no.~8, 2019.

\bibitem{keating1966}
\BIBentryALTinterwordspacing
P.~N. Keating, ``Effect of invariance requirements on the elastic strain energy of crystals with application to the diamond structure,'' \emph{Phys. Rev.}, vol. 145, pp. 637--645, May 1966. [Online]. Available: \url{https://link.aps.org/doi/10.1103/PhysRev.145.637}
\BIBentrySTDinterwordspacing

\bibitem{ager1991spatially}
J.~W. Ager~III, D.~K. Veirs, and G.~M. Rosenblatt, ``Spatially resolved raman studies of diamond films grown by chemical vapor deposition,'' \emph{Physical Review B}, vol.~43, no.~8, p. 6491, 1991.

\bibitem{yoshikawa1995raman}
M.~Yoshikawa, Y.~Mori, H.~Obata, M.~Maegawa, G.~Katagiri, H.~Ishida, and A.~Ishitani, ``Raman scattering from nanometer-sized diamond,'' \emph{Applied Physics Letters}, vol.~67, no.~5, pp. 694--696, 1995.

\bibitem{chaigneau2012laser}
M.~Chaigneau, G.~Picardi, H.~A. Girard, J.-C. Arnault, and R.~Ossikovski, ``Laser heating versus phonon confinement effect in the raman spectra of diamond nanoparticles,'' \emph{Journal of Nanoparticle Research}, vol.~14, pp. 1--8, 2012.

\bibitem{aoki1980}
\BIBentryALTinterwordspacing
H.~Aoki, ``Real-space renormalisation-group theory for anderson localisation: decimation method for electron systems,'' \emph{Journal of Physics C: Solid State Physics}, vol.~13, no.~18, p. 3369, jun 1980. [Online]. Available: \url{https://dx.doi.org/10.1088/0022-3719/13/18/006}
\BIBentrySTDinterwordspacing

\bibitem{vanoni2023renormalization}
C.~Vanoni, B.~L. Altshuler, V.~E. Kravtsov, and A.~Scardicchio, ``Renormalization group analysis of the anderson model on random regular graphs,'' \emph{arXiv preprint arXiv:2306.14965}, 2023.

\bibitem{thouless1974electrons}
D.~J. Thouless, ``Electrons in disordered systems and the theory of localization,'' \emph{Physics Reports}, vol.~13, no.~3, pp. 93--142, 1974.

\bibitem{nakayama1994dynamical}
T.~Nakayama, K.~Yakubo, and R.~L. Orbach, ``Dynamical properties of fractal networks: Scaling, numerical simulations, and physical realizations,'' \emph{Reviews of modern physics}, vol.~66, no.~2, p. 381, 1994.

\bibitem{meakin1984effects}
P.~Meakin, ``Effects of cluster trajectories on cluster-cluster aggregation: A comparison of linear and brownian trajectories in two-and three-dimensional simulations,'' \emph{Physical Review A}, vol.~29, no.~2, p. 997, 1984.

\bibitem{brasil1999recipe}
A.~M. Brasil, T.~L. Farias, and M.~G. Carvalho, ``A recipe for image characterization of fractal-like aggregates,'' \emph{Journal of Aerosol Science}, vol.~30, no.~10, pp. 1379--1389, 1999.

\bibitem{brasil2000evaluation}
A.~Brasil, T.~Farias, and M.~Carvalho, ``Evaluation of the fractal properties of cluster-cluster aggregates,'' \emph{Aerosol Science \& Technology}, vol.~33, no.~5, pp. 440--454, 2000.

\bibitem{brasil2001numerical}
A.~Brasil, T.~L. Farias, M.~d.~G. Carvalho, and {\"U}.~{\"O}. K{\"o}yl{\"u}, ``Numerical characterization of the morphology of aggregated particles,'' \emph{Journal of Aerosol Science}, vol.~32, no.~4, pp. 489--508, 2001.

\bibitem{Mathematica}
\BIBentryALTinterwordspacing
W.~R. Inc., ``Mathematica, {V}ersion 13.1,'' champaign, IL, 2022. [Online]. Available: \url{https://www.wolfram.com/mathematica}
\BIBentrySTDinterwordspacing

\bibitem{rammal1983nature}
R.~Rammal, ``Nature of eigenstates on fractal structures,'' \emph{Physical Review B}, vol.~28, no.~8, p. 4871, 1983.

\bibitem{wang1995localization}
X.~R. Wang, ``Localization in fractal spaces: Exact results on the sierpinski gasket,'' \emph{Physical Review B}, vol.~51, no.~14, p. 9310, 1995.

\bibitem{pal2012}
\BIBentryALTinterwordspacing
B.~Pal and A.~Chakrabarti, ``Staggered and extreme localization of electron states in fractal space,'' \emph{Phys. Rev. B}, vol.~85, p. 214203, Jun 2012. [Online]. Available: \url{https://link.aps.org/doi/10.1103/PhysRevB.85.214203}
\BIBentrySTDinterwordspacing

\end{thebibliography}
\end{document}